\documentclass{article}

\PassOptionsToPackage{numbers, compress}{natbib}

\usepackage[preprint]{neurips_2020}

\usepackage[utf8]{inputenc} 
\usepackage[T1]{fontenc}    
\usepackage{hyperref}       
\usepackage{url}            
\usepackage{booktabs}       
\usepackage{amsfonts}       
\usepackage{nicefrac}       
\usepackage{microtype}      
\usepackage{enumitem}
\usepackage{graphicx}
\usepackage{makecell}
\usepackage{algorithm}
\usepackage{algorithmic}

\usepackage{amsmath,amsfonts,bm}
\allowdisplaybreaks
\newcommand\Tstrut{\rule{0pt}{2.3ex}}
\newcommand\Bstrut{\rule[-0.9ex]{0pt}{0pt}}

\def\eqref#1{equation~\ref{#1}}

\def\1{\bm{1}}

\DeclareMathAlphabet{\mathsfit}{\encodingdefault}{\sfdefault}{m}{sl}
\SetMathAlphabet{\mathsfit}{bold}{\encodingdefault}{\sfdefault}{bx}{n}

\title{Adaptive Invariance for Molecule Property Prediction}

\author{%
  Wengong Jin $\quad$ Regina Barzilay $\quad$ Tommi Jaakkola \\
  CSAIL, Masssachusetts Institute of Technology \\
  \texttt{\{wengong,regina,tommi\}@csail.mit.edu}
}

\begin{document}

\maketitle
\begin{abstract}
    Effective property prediction methods can help accelerate the search for COVID-19 antivirals either through accurate in-silico screens or by effectively guiding on-going at-scale experimental efforts. However, existing prediction tools have limited ability to accommodate scarce or fragmented training data currently available. In this paper, we introduce a novel approach to learn predictors that can generalize or extrapolate beyond the heterogeneous data. Our method builds on and extends recently proposed invariant risk minimization, adaptively forcing the predictor to avoid nuisance variation. We achieve this by continually exercising and manipulating latent representations of molecules to highlight undesirable variation to the predictor. To test the method we use a combination of three data sources: SARS-CoV-2 antiviral screening data, molecular fragments that bind to SARS-CoV-2 main protease and large screening data for SARS-CoV-1. Our predictor outperforms state-of-the-art transfer learning methods by significant margin. We also report the top 20 predictions of our model on Broad drug repurposing hub.

\end{abstract}

\section{Introduction}

The race to identify promising repurposing drug candidates against COVID-19 calls for improvements in the underlying property prediction methodology. The accuracy of many existing techniques depends heavily on access to reasonably large, uniform training data. Such high-throughput, on target screening data is not yet publicly available for COVID-19. Indeed, we have only 48 drugs with measured in-vitro SARS-CoV-2 activity shared with the research community~\cite{jeon2020identification}. This limited data scenario is not unique to the current pandemic but likely to recur with each evolving or new viral challenge. The ability to make accurate predictions based on all the available data, however limited, is also helpful in guiding later high-throughput targeted experimental effort.

We can supplement scarce on-target data with other related data sources, either related screens pertaining to COVID-19 or screens involving related viruses. For instance, we can use additional data pertaining to molecular fragment screens that measure binding to SARS-CoV-2 main protease, obtained via crystallography screening~\cite{diamond}. On average, these fragments consist of only 14 atoms, comprising roughly 37\% of full drug size molecules. Another source of data is SARS-CoV-1 screens. Since SARS-CoV-1 and SARS-CoV-2 proteases are similar (more than 79\% sequence identity)~\cite{zhou2020pneumonia}, drugs screened against SARS-CoV-1 can be expected to be relevant for SARS-CoV-2 predictions. These two examples highlight the challenges for property prediction tools: much of the available training data comes from either different chemical space (molecular fragments) or different viral species (SARS-CoV-1).

The key technological challenge is to be able to estimate models that can extrapolate beyond their training data, e.g., to different chemical spaces. The ability to extrapolate implies a notion of invariance (being impervious) to the differences between the available training data and where predictions are sought. A recently proposed approach known as invariant risk minimization (IRM)~\cite{arjovsky2019invariant} seeks to find predictors that are simultaneously optimal across different such scenarios (called environments). Indeed, the differences in chemical spaces can be thought as "nuisance variation" that the predictor should be explicitly forced to ignore. One possible way to automatically define this type of environment variability for molecules is scaffolds~\cite{bemis1996properties}. But the setting is challenging since scaffolds are combinatorial descriptors (substructures) and can potentially uniquely identify each compound in the training data. Useful environments for estimation should enjoy some statistical support.

In this paper we propose a novel variant of invariant risk minimization specifically tailored to rich, combinatorially defined environments typical in molecular contexts. Indeed, unlike in standard IRM, we introduce two dynamic (in contrast to many static) environments. These are defined over the same set of training examples, but differ in terms of their associated latent representations. The difference between them arises from continually adjusted perturbations that manipulate the latent representations of compounds towards more ``generic'' versions with the help of a scaffold classifier. The idea is to explicitly highlight to the property predictor that operates on these latent representations what the nuisance variability is that it should not rely on.

Our method is evaluated on existing SARS-CoV-2 screening data~\cite{diamond,jeon2020identification}. The training utilizes three sources of data: SARS-CoV-2 screened molecules, SARS-CoV-2 fragments and SARS-CoV-1 screening data described above. 
We compare against multiple transfer learning techniques such as domain adversarial training~\citep{ganin2016domain} and conditional domain adversarial network~\citep{long2018conditional}. On two SARS-CoV-2 datasets, the proposed approach outperforms the best performing baseline with 8-16\% relative AUROC improvement. Finally, we apply our model on Broad drug repurposing hub~\cite{corsello2017drug} and report the top 20 predictions for further investigation.

\section{Domain Extrapolation}
Training data in many emerging applications is necessarily limited, fragmented, or otherwise heterogeneous. It is therefore important to ensure that model predictions derived from such data generalize substantially beyond where the training samples lie. In other words, the trained model should have the ability to extrapolate. For instance, in computational chemistry, it is desirable for property prediction models to perform well in time-split scenarios where the evaluation concerns compounds that were created after those in the training set. Another way to simulate  evaluation on future compounds is through a scaffold split~\cite{bemis1996properties}. A scaffold split between training and test introduces some structural separation between the chemical spaces of the two sets of compounds, hence evaluating the model's ability to extrapolate to a new domain. 

One way to ensure domain extrapolation is to enforce an appropriate invariance criterion during training. We envision here that the compounds $X$ can be divided into potentially a large number of domains or ``environments'' $E$, for example, based on their scaffold. The goal is then to learn a parametric mapping $Z=\phi(X)$ of compounds $X$ to their latent representations $Z$ in a manner that satisfies the chosen invariance criterion.  A number of such strategies relevant to extrapolation have been proposed. They can be roughly divided into the following three categories: 
\begin{itemize}[leftmargin=*,topsep=0pt,itemsep=0pt]
    \item Domain adversarial training~\cite{ganin2016domain} enforces the latent representation $Z = \phi(X)$ to have the same distribution across different domains $E$. If we denote by $P(X|E)$ the conditional distribution of compounds in environment $E$, then we want $P(\phi(X)|E) = P(\phi(X))$ for all $E$. With some abuse of notation, we can write this condition as $Z \perp E$. A single predictor is learned based on $Z=\phi(X)$, i.e.,  all the domains share the same predictor. As a result, the predicted label distribution $P(Y)$ will also be the same across the domains. This can be problematic when the training and test domains have very different label distributions~\cite{zhao2019learning}. The independence condition itself can be challenging to satisfy when the chemical spaces overlap across the environments. 
    
    \item Conditional domain adaptation~\cite{long2018conditional} relaxes the requirement that the label distributions must agree across the environments. The key idea is to condition the invariance criterion on the label. In other words, we require that $P(\phi(X)|E,Y) = P(\phi(X)|Y)$ for all $E$ and $Y$, i.e., we aim to satisfy the independence statement $Z \perp E \;|\; Y$. The formulation allows the label distribution to vary between domains since $E$ and $Y$ can depend on each other. The constraint remains, however, too restrictive about the latent representation. To illustrate this, consider a simple case where the environments share the same chemical space and differ only in terms of proportions of different types of compounds in them. These type proportions play roles analogous to label proportions in domain adversarial training. Hence, the only way to achieve $Z \perp E \;|\; Y$ would be if the proportions were the same across environments. To state the example differently, a functional mapping $Z=\phi(X)$ cannot fractionally assign probability mass placed on $X$ to different latent space locations $Z$; it all has to be mapped to a single location. To reduce the impact of the strict condition, we would have to introduce $P(Z|\phi(X),E)$ in place of the simpler functional mapping $Z=\phi(X)$, further complicating the approach.
    
    \item Invariant risk minimization (IRM)~\cite{arjovsky2019invariant} seeks a different notion of invariance, focusing less on aligning distributions of latent representations, and instead shifting the emphasis on how those representations can be consistently used for predictions. The IRM principle requires that the predictor $f$ operating on $Z=\phi(X)$ is simultaneously optimal across different environments or domains. For example, this holds if our representation explicates only features that are (causally) necessary for the correct prediction. How $\phi(X)$ is distributed across the environments is then immaterial. The associated conditional independence criterion is $Y \perp E \;\vert\; Z$. In other words, knowing the environment shouldn't provide any additional information about $Y$ beyond the features $Z=\phi(X)$. The distribution of labels can differ across the environments. 
\end{itemize}

While the IRM principle provides a natural framework for domain extrapolation, it needs to be extended in several ways for our setting. The main limitation of the original framework is that the environments $E$ themselves are fixed and pre-defined. Their role in the pricinple is to illustrate ``nuisance'' variation, i.e., variability that the predictor should learn not to rely on. In order to enforce the associated independence criterion, we need a fair number of examples within each such environment. The approach therefore becomes unsuitable when the natural environments such as scaffolds are combinatorially defined or otherwise have high cardinality. Indeed, we might have only a single molecule per environment in our training set, making the independence criterion vacuous ($E$ would uniquely specify $X$, thus also $Z$ and $Y$). A straightforward remedy for the high cardinality environments would be to introduce a coarser definition, and enforce the principle at this coarse level instead. Since environments represent constraints on the predictor, their role in estimation is adversarial. What is then the appropriate trade-off between such a coarser definition (relaxation of constraints) and our  ability to predict? We side-step having to answer this question, and instead propose to dynamically map the large number of environments to just two. These two environments are designed to nevertheless highlight the nuisance variation the predictor should avoid but do so in a tractable manner.  

\section{IRM with adaptive environments}

\begin{figure}[t]
    \centering
    \includegraphics[width=\textwidth]{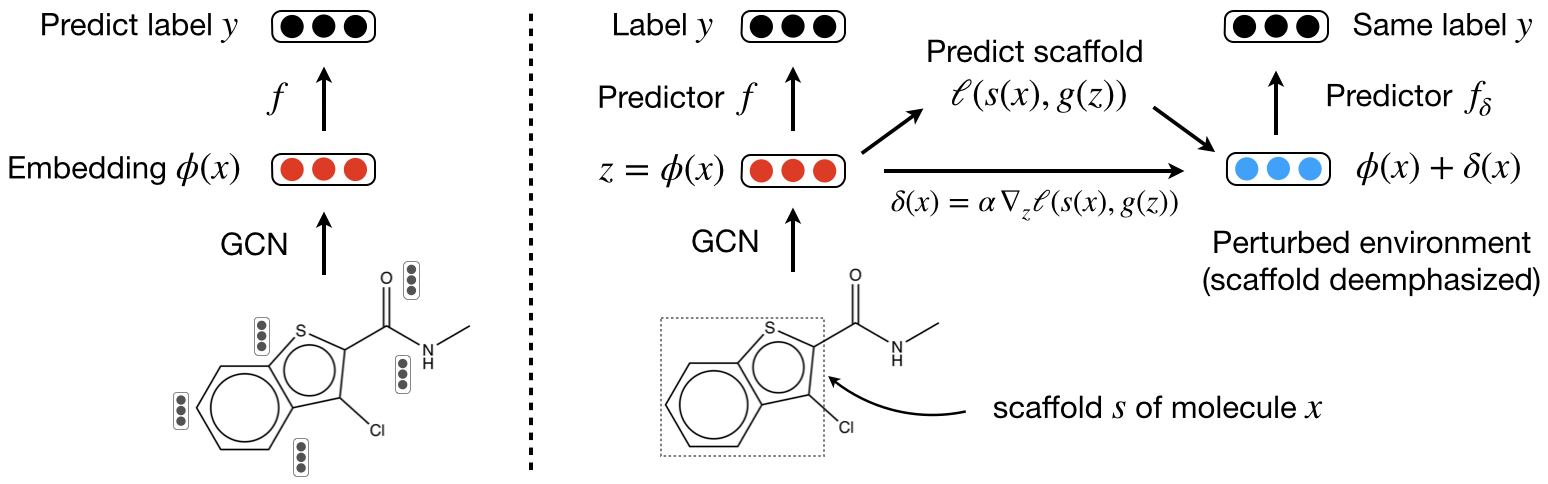}
    \caption{\textbf{Left}: Illustration of base GCN model for molecule property prediction. \textbf{Right}: IRM with adaptive environments. The model perturbs the representation $\phi(x)$ into $\phi(x) + \delta(x)$ using the gradient from scaffold classifier $g$. The predictor $f$ is trained to be simultaneously optimal across two environments. The scaffold $s$ is a subgraph of a molecule $x$ with its side chains removed~\cite{bemis1996properties}.}
    \label{fig:chemprop}
\end{figure}

Our goal is to adaptively highlight to the predictor the type of variability that it ought not to rely on. We do this by replacing high cardinality environments such as those based on scaffold with just two new environments. These two new environments are unusual in the sense that they share the exact same set of examples. Indeed, they only differ in terms of the representation that the predictor operates on. The first environment simply corresponds to the representation we are trying to learn, i.e., $z=\phi(x)$, where the lowercase letters refer to specific instances rather than random variables. The second environment is defined in terms of a modified representation $\phi(x)+\delta(x)$ that is a perturbed version of $\phi(x)$ and constructed with the help of the environment or scaffold classifier. The goal of $\delta(x)$ is to explicate directions in the latent representation that the predictor should avoid paying attention to. While traditional IRM environments divide examples $x$ into environments, often exclusively, we instead exercise different latent representations over the same set of examples. 

More formally, our two environments correspond to a choice of perturbation $h\in \{0,\delta\}$ used to derive the latent representation $z$ from $x$, i.e., $\phi(x)+h(x)$. The associated target labels are clearly the same regardless of which perturbation (none or $\delta(x)$) was chosen. The key part of our approach pertains to how $\delta(x)$ is defined. To this end, let $g(\phi(x)))$ be a parametric environment classifier that we will instantiate in detail later. The associated classification loss is $\ell(s(x), g(\phi(x)))$ where $s(x)$ is the correct original environment label (here a scaffold) for $x$. The scaffold classifier is evolved together with the feature mapping $\phi$ and the associated predictor $f$. We define the non-zero perturbation $\delta(x)$ in terms of the gradient: 
\begin{equation}
    \delta(x) = \alpha \nabla_{z} \ell(s(x), g(z))_{|z=\phi(x)}
\end{equation}
where $\alpha$ is a step size parameter. The goal of this perturbation is to turn $\phi(x)$ into its ``generic'' version $\phi(x)+\delta(x)$ which contains less information of the environment (e.g., scaffold). Note that if we were to perform adversarial domain alignment, $\delta(x)$ would represent a reverse gradient update to modify $\phi(x)$. We do not do that, instead we are using the perturbation to highlight directions of variability to avoid for the predictor $f$ within an overall IRM formulation. The degree to which $\phi(x)$ is adjusted in response to $\delta(x)$ arises from the IRM principle, not from a direct alignment objective. 

We begin by building the overall training objective which is then optimized in batches as described in Algorithm 1.  Let $(x_i,y_i)$ be a pair of training example + the associated label to predict. Each $x_i$ also has an environment label/features given by $s_i = s(x_i)$ (the original mapping of examples to environments is assumed given and fixed, defined by $s(x)$). The environment classifier is trained to minimize 
\begin{equation}
    \mathcal{L}^e(g \circ \phi) = \sum_i \ell(s_i, g(\phi(x_i)))
\end{equation}
As we will explain later on, the environment classifier remains ``unaware'' of how the perturbation is derived on the basis of its predictions. The loss of the predictor $f$, now operating on $\phi(x)+h(x)$, where $h\in \{0,\delta\}$, is defined as 
\begin{equation}
    \mathcal{L}(f\circ (\phi+h)) = \sum_i \ell\big(y_i, f(\phi(x_i)+h(x_i))\big) \qquad h\in \{0, \delta\}
\end{equation}
The specific form of the loss depends on the prediction task. 
In accordance with the IRM principle, we enforce that the predictor operating on $z=\phi(x)+h(x)$ remains optimal whether its input is $\phi(x)$ or the perturbed version $\phi(x)+\delta(x)$. In other words, we require that 
\begin{equation}
\mathcal{L}(f\circ (\phi+h))\leq \min_{h\in \{0,\delta\}} \mathcal{L}(f_h\circ (\phi+h))
\end{equation}
where $f_h$ is a predictor in the same parametric family as $f$ but trained separately with the knowledge of $h$ (perturbed or not). By relaxing the constraints via Lagrange multipliers, we express the overall training objective as 
\begin{equation}
    \lambda_e \mathcal{L}^e(g \circ \phi) + \sum_{h \in \{0,\delta\}} \bigg[ 
    (1+\lambda_h)\mathcal{L}(f \circ (\phi+h)) - \lambda_h 
    \mathcal{L}(f_h \circ (\phi+h))\bigg]
    \label{eq:cirm}
\end{equation}
This minimax objective is \emph{minimized} with respect to $\phi$, $g$, and $f$, and \emph{maximized} with respect to $f_h$, $h\in\{0,\delta\}$. A few remarks are necessary concerning this objective:
\begin{itemize}[leftmargin=*,topsep=0pt,itemsep=0pt]
    \item Even though $\delta$ is defined on the basis of $\phi$ and the environment classifier $g$, we view it as a functionally independent player. The goal of $\delta$ is to enforce optimality of $f$ and therefore it plays an adversarial role relative to $f$. Similarly to GAN objectives where the discriminator has a separate objective function, different from the generator, we separate out $\delta$ as another player in an overall game theoretic objective. Specifically, $\delta$ takes input from $\phi$ and $g$ but does not inform them in return in back-propagation.\footnote{Incorporating this higher order dependence would not improve the empirical results.}
    \item $\phi$ in our objective is adjusted to also help the auxiliary environment classifier. This is contrary to domain alignment where the goal would be to take out any dependence on the environment. The benefit in our formulation is two-fold. First, the term grounds $\phi$ also based on the auxiliary objective, helping it to retain useful information about each example $x$. Second, the term grounds and stabilizes the definition of $\delta$ as the gradient of the environment predictor since $g$ no longer approaches a random predictor. It would be weak if $\phi$ contains no information about the environment as in domain alignment. Thus $\delta$ remains well-defined as a direction throughout the optimization.
\end{itemize}   
The training procedure is shown in Algorithm 1.

\begin{algorithm}[t]
\label{alg:cirm}
\begin{algorithmic}[1]
\caption{Training IRM with adaptative environments}
\FOR{each training step}
\STATE Sample a batch of molecules $\{x_1,\cdots,x_n\}$ with their environments $\{s_1,\cdots,s_n\}$
\STATE Encode molecules $\{x_1,\cdots,x_n\}$ into vectors $\{\phi(x_1),\cdots,\phi(x_n)\}$
\STATE Computed environment classification loss $\mathcal{L}^s(g \circ \phi) = \sum_i \ell(s_i, g(\phi(x_i)))$.
\STATE Construct perturbed representation $\delta(x_i) = -\alpha \nabla_{\phi(x_i)} \ell(s_i, g(\phi(x_i)))$
\STATE Compute invariant predictor loss $\mathcal{L}(f \circ \phi)$
\STATE Compute competing predictor loss $\mathcal{L}(f_h \circ (\phi + h)) \quad h \in \{0, \delta\}$
\STATE Update $\phi, f, g$ to minimize Eq.~(\ref{eq:cirm})
\STATE Update $f_0, f_\delta$ to maximize Eq.~(\ref{eq:cirm})
\ENDFOR
\end{algorithmic}
\end{algorithm}

\subsection{Adapting the framework to molecule property prediction }
In molecule property prediction, the training data is a collection of pairs $\{(x_i,y_i)\}$, where $x_i$ is a molecular graph and $y_i$ is its activity score, typically binary (active/inactive).
The feature extractor $\phi(\cdot)$ is a graph convolutional network (GCN) which translates a molecular graph into a continuous vector through directed message passing operations~\cite{yang2019analyzing}. The predictor $f$ is a feed-forward network that takes $\phi(x)$ or $\phi(x)+\delta(x)$ as input and yields predicted activity $y$. 

The original environment of each compound $x_i$ is defined as its Murcko scaffold~\cite{bemis1996properties}, which is a subgraph of $x_i$. Since scaffold is a combinatorial object with a large vocabulary of possible values, we define and train the environment classifier in a contrastive fashion~\cite{oord2018representation}. 
Specifically, for a given molecule $x_i$ with scaffold $s_i$, we randomly sample $n$ other molecules and take their associated scaffolds $\{s_k\}$ as negative examples, as the contrastive set $C$. The environment classifier $g$ makes use of a feed-forward network $g_s$ that maps each compound or a scaffold (subgraph) to a feature vector. The probability that $x_i$ is mapped to its correct scaffold $s_i$ is then defined as
\begin{equation}
P_g(s=s_i|x_i,C) = 
\frac{\exp\{\mathrm{sim}\big(g_s(\phi({x_i})), g_s(\phi(s_i))\big)\}}{\sum_{k\in \{i\}\cup C} \exp\{\mathrm{sim}\big(g_s(\phi(x_i)), g_s(\phi(s_k))\big)\}}
\label{eq:scaf}
\end{equation}
where $\mathrm{sim(\cdot,\cdot)}$ stands for cosine similarity. In practice, we use the molecules within the same batch as negative examples. 

\section{Experiments}

Our experiments consist of two settings. To compare our method with existing transfer learning techniques, we first evaluate our methods on a standard unsupervised transfer setup. All the models are trained on SARS-CoV-1 data and tested on SARS-CoV-2 compounds. Next, in order to identify drug candidates for SARS-CoV-2, we extend our method by incorporating labeled SARS-CoV-2 data to maximize prediction accuracy and perform virtual screening over Broad drug repurposing hub~\cite{corsello2017drug}.

\paragraph{Training data} Our training data consist of three screens related to SARS-CoV. All the data can be found at \url{https://github.com/yangkevin2/coronavirus_data}.

\begin{itemize}[leftmargin=*,topsep=0pt,itemsep=0pt]
    \item \textbf{SARS-CoV-2 MPro inhibition } 881 fragments screened for SARS-CoV-2 main protease (Mpro) collected by the Diamond Light Source group~\cite{diamond}. The dataset contains 78 hits. 
    \item \textbf{SARS-CoV-2 antiviral activity } 48 FDA-approved drugs screened for antiviral activity against SARS-CoV-2 in vitro~\cite{jeon2020identification}, including reference drugs such as Remdesivir, Lopinavir and Chloroquine. The dataset contains 27 hits.
    \item \textbf{SARS-CoV-1 3CLpro inhibition } Over 290K molecules screened for activity against SARS-CoV-1 3C-like protease (3CLpro) in PubChem AID1706 assay. There are 405 active compounds. 
\end{itemize}

\paragraph{Baselines } We compare the proposed approach with the following baselines:
\begin{itemize}[leftmargin=*,topsep=0pt,itemsep=0pt]
    \item \textbf{Direct transfer}: We train a GCN on SARS-CoV-1 data and directly test it on SARS-CoV-2 data.
    
    \item \textbf{Domain adversarial training} (DANN)~\cite{ganin2016domain}: Since distribution of molecules is different between SARS-CoV-1 and SARS-CoV-2 datasets, we use domain adversarial training to facilitate transfer. Specifically, we augment our GCN with additional domain classifier $g$ to enforce the distribution of $\phi(x)$ to be the same across training (SARS-CoV-1) and test set (SARS-CoV-2).
    
    \item \textbf{Conditional adversarial domain adaptation} (CDAN)~\cite{long2018conditional} conditions the domain classifier $g$ with predicted labels $f(\phi(x))$. In particular, we adopt their multilinear conditioning strategy: the input to $g$ becomes a vector outer-product $\phi(x) \otimes f(\phi(x))$, which has the same dimension as $\phi(x)$ for binary classification tasks.
    
    \item \textbf{Scaffold adversarial training} (SANN): This is an extension of DANN where the domain classifier $g$ is replaced with our scaffold classifier $g_s$ in  Eq.(\ref{eq:scaf}). SANN seeks to learn a scaffold-invariant representation $\phi(x)$ through the following minimax game ($\mathcal{L}^e$ is scaffold classification loss):
    \begin{equation}
        \min_{\phi,f} \max_g \mathcal{L}(f \circ \phi) - \lambda_e \mathcal{L}^e(g_s \circ \phi)
    \end{equation}

    \item \textbf{Invariant risk minimization} (IRM): The original IRM~\cite{arjovsky2019invariant} requires the predictor $f$ to be constant, which does not work well in our setting. Therefore, we adopt an adversarial formulation for IRM proposed in \cite{chang2020invariant}, allowing us to use powerful neural predictors:
    \begin{equation}
        \min_{\phi,f} \max_{f_1,\cdots,f_K} \sum_{e} (1 + \lambda_e) \mathcal{L}(f \circ \phi) - \lambda_e \mathcal{L}(f_e \circ \phi)
    \end{equation}
    Here each of the $K$ environments consists of molecules with the same scaffold. Since the number of environments is large, we impose parameter sharing among the competing predictors $f_1,\cdots,f_K$. Specifically, the input of $f_j=f'(\phi(x), j)$ is a concatenation of $\phi(x)$ and one-hot encoding of $j$.
\end{itemize}

\paragraph{Model hyperparameters } For our model, we set $\lambda_h = \lambda_e = 0.1$ and perturbation learning rate $\alpha=0.1$, which worked well across all experiments. All methods are trained with Adam using its default configuration. Our GCN implementation is based on chemprop~\cite{yang2019analyzing}. 
\begin{itemize}[leftmargin=*,topsep=0pt,itemsep=0pt]
    \item For unsupervised transfer, we use their default hyper-parameter setting. For all methods, the GCN $\phi$ has three layers with hidden layer dimension 300. The predictor $f$ is a two-layer MLP. 
    \item For supervised transfer, we perform hyper-parameter optimization to identify the best architecture for the multitask GCN. The GCN $\phi$ has two layers with hidden layer dimension 2000. The predictor $f$ is a three-layer MLP. The dropout rate is $0.1$. For fair comparison, all the methods use the same architecture in this setting.
\end{itemize}

\subsection{Unsupervised transfer}
\textbf{Setup } Our model is a single-task binary classification model which predicts the SARS-CoV-1 3CLpro inhibition. After training, the model is tested on SARS-CoV-2 Mpro and antiviral data. Each model is evaluated under five independent runs and we report the average AUROC score. 

\textbf{Results } Our results are shown in Table~\ref{tab:unsup}. The proposed method significantly outperformed all the baselines, especially on the Mpro inhibition prediction dataset (0.756 versus 0.653 AUROC). 

\textbf{Ablation study } Indeed, the improvement of our model comes from two sources: the additional auxiliary task and IRM principle. To show individual contribution of each component, we conduct an ablation study of our method without the IRM principle. The loss function in this case is the scaffold classification loss plus property prediction loss $\lambda_e \mathcal{L}^e(g\circ \phi) + \mathcal{L}(f \circ \phi)$. The performance of this method is shown in the end of Table~\ref{tab:unsup} (``without IRM''). The auxiliary scaffold classifier shows quite significant improvement, but is still inferior to our full model trained with IRM principle.

\begin{table}[t]
\centering
\caption{Unsupervised transfer results. Models are trained on SARS-CoV-1 data and tested on SARS-CoV-2 compounds. We report average of AUROC under 5-fold cross validation.}
\begin{tabular}{lcc}
\hline
 & Mpro inhibition AUC & Antiviral activity AUC \Tstrut\Bstrut \\
\hline
Direct Transfer & $0.642 \pm 0.021$ & $0.415 \pm 0.081$ \Tstrut\Bstrut \\
DANN~\cite{ganin2016domain} & $0.646 \pm 0.012$ & $0.607 \pm 0.121$  \Tstrut\Bstrut \\
SANN  & $0.630 \pm 0.079$ & $0.570 \pm 0.096$  \Tstrut\Bstrut \\
CDAN~\cite{long2018conditional} & $0.625 \pm 0.013$ & $0.639 \pm 0.067$  \Tstrut\Bstrut \\
IRM~\cite{arjovsky2019invariant} & $0.653 \pm 0.022$ &  $0.391 \pm 0.086$ \Tstrut\Bstrut \\
\hline
Our method & $\mathbf{0.756 \pm 0.012}$ & $\mathbf{0.695 \pm 0.068}$  \Tstrut\Bstrut \\
 - without IRM & $0.736 \pm 0.007$ & $0.678 \pm 0.057$ \Tstrut\Bstrut \\
\hline
\end{tabular}
\label{tab:unsup}
\end{table}

\subsection{Drug repurposing for SARS-CoV-2}
\textbf{Setup }  We extend all the methods to multitask binary classification models that predict \textit{three} different properties for each new compound: 1) probability of inhibiting the SARS-CoV-2 Mpro; 2) antiviral activity against SARS-CoV-2; 3) probability of inhibiting SARS-CoV-1 3CLpro. 

Each model is evaluated under 5-fold cross validation with the same splits. In each fold, the training set contains the SARS-CoV data and 60\% of the SARS-CoV-2 data (Mpro + antiviral), and the test set contains the rest 40\% of the SARS-CoV-2 compounds. We report the mean and standard deviation of AUROC score evaluated on five different folds.

\textbf{Results } Our results are shown in Table~\ref{tab:sup}. The proposed method significantly outperformed the two baselines, especially on the antiviral activity prediction dataset (0.89 versus 0.82 AUROC).
As an ablation study, we also trained a GCN on only SARS-CoV-2 data (the first row in Table~\ref{tab:sup}). Indeed, the multitask GCN trained with additional SARS-CoV-1 data performs better (0.740 vs 0.807 on antiviral prediction), indicating that the two virus are closely related.

\begin{table}[t]
\centering
\caption{5-fold cross validation of different methods. CoV-2 means that the model is trained on SARS-CoV-2 data. CoV-1 means that the model is additionally trained on SARS-CoV-1 data.}
\begin{tabular}{lcccc}
\hline
Method & CoV-1 & CoV-2 & Mpro inhibition AUC & Antiviral activity AUC \Tstrut\Bstrut \\
\hline
Multitask GCN & & $\checkmark$ & $0.7646 \pm 0.0398$ & $0.7404 \pm 0.1117$ \Tstrut\Bstrut \\
Multitask GCN & $\checkmark$ & $\checkmark$ & $0.7841 \pm 0.0416$ & $0.8067 \pm 0.0690$ \Tstrut\Bstrut \\
DANN & $\checkmark$ & $\checkmark$ & $0.7785 \pm 0.0321$ & $0.8146 \pm 0.1068$ \Tstrut\Bstrut \\
IRM~\cite{arjovsky2019invariant} & $\checkmark$ & $\checkmark$ & $0.7778 \pm 0.0426$ & $0.8198 \pm 0.1072$ \Tstrut\Bstrut \\
\hline
Our method & $\checkmark$ & $\checkmark$ & $\mathbf{0.7955 \pm 0.0489}$ & $\mathbf{0.8930 \pm 0.0267}$ \Tstrut\Bstrut \\
\hline
\end{tabular}
\label{tab:sup}
\end{table}

The best model is then used to predict the SARS-CoV-2 Mpro inhibition and antiviral activity of compounds in Broad drug repurposing hub. In order to utilize maximal amount of labeled data, the model is re-trained under 10-fold cross validation with 90\%/10\% split (instead of 60\%/40\%). The resulting 10 models are combined together as an ensemble to predict properties for new compounds. 
We report the top 20 predicted molecules for MPro inhibition and antiviral activity in Table~\ref{tab:mpro} and \ref{tab:fda}.

\begin{table}[t]
\small
\centering
\caption{Top 20 SARS-CoV-2 Mpro inhibiting molecules (non-covalent) in the Broad drug repurposing hub.}
\begin{tabular}{lc}
\hline
SMILES & MPro \Tstrut\Bstrut \\
\hline
Nc1ccc(cc1)S(N)(=O)=O & 0.797 \Tstrut\Bstrut \\
N\#Cc1ccncn1 & 0.769 \Tstrut\Bstrut \\
Nc1ccccc1S(N)(=O)=O & 0.747 \Tstrut\Bstrut \\
Cc1ccc(cc1)S(N)(=O)=O & 0.727 \Tstrut\Bstrut \\
NS(=O)(=O)c1cc(Cl)c(Cl)c(c1)S(N)(=O)=O & 0.695 \Tstrut\Bstrut \\
NCc1ccc(cc1)S(N)(=O)=O & 0.694 \Tstrut\Bstrut \\
NC(=N)NCCNS(=O)(=O)c1cccc2cnccc12 & 0.679 \Tstrut\Bstrut \\
NC(=N)NS(=O)(=O)c1ccc(N)cc1 & 0.662 \Tstrut\Bstrut \\
NS(=O)(=O)NCc1csc2ccccc12 & 0.651 \Tstrut\Bstrut \\
NS(=O)(=O)C\#Cc1ccccc1 & 0.614 \Tstrut\Bstrut \\
Nc1cc(C(Cl)=C(Cl)Cl)c(cc1S(N)(=O)=O)S(N)(=O)=O & 0.613 \Tstrut\Bstrut \\
NC(=N)NCC1COC2(CCCCC2)O1 & 0.582 \Tstrut\Bstrut \\
NC(=N)NC(=O)c1cnccn1 & 0.580 \Tstrut\Bstrut \\
NC(=N)Nc1ccc(cc1)C(=O)Oc1ccc2cc(ccc2c1)C(N)=N & 0.560 \Tstrut\Bstrut \\
N\#Cc1ncccn1 & 0.524 \Tstrut\Bstrut \\
Nc1ccncc1N & 0.521 \Tstrut\Bstrut \\
N\#Cc1ccc(cc1)C1CCCc2cncn12 & 0.450 \Tstrut\Bstrut \\
NS(=O)(=O)Cc1noc2ccccc12 & 0.443 \Tstrut\Bstrut \\
NS(=O)(=O)c1ccc(NC(=O)Nc2ccc(F)cc2)cc1 & 0.442 \Tstrut\Bstrut \\
\hline
\end{tabular}
\label{tab:mpro}
\end{table}

\begin{table}[t]
\scriptsize
\centering
\caption{Top 20 SARS-CoV-2 antiviral molecules in the Broad drug repurposing hub.}
\begin{tabular}{lc}
\hline
SMILES & Antiviral \Tstrut\Bstrut \\
\hline
C[C@]12CC(=O)[C@H]3[C@@H](CCC4=CC(=O)CC[C@]34C)[C@@H]1CC[C@]2(O)C(=O)CO & 0.955 \Tstrut\Bstrut \\
C[C@]1(O)CC[C@H]2[C@@H]3CCC4=CC(=O)CC[C@]4(C)[C@H]3CC[C@]12C & 0.953 \Tstrut\Bstrut \\
C[C@]12C[C@H](O)[C@H]3[C@@H](CCC4=CC(=O)CC[C@]34C)[C@@H]1CC[C@]2(O)C(=O)CO & 0.948 \Tstrut\Bstrut \\
CC(=O)OCC(=O)[C@@]1(O)CC[C@H]2[C@@H]3CCC4=CC(=O)CC[C@]4(C)[C@H]3C(=O)C[C@]12C & 0.945 \Tstrut\Bstrut \\
C[C@H]1C[C@H]2[C@@H]3CC[C@](O)(C(C)=O)[C@@]3(C)CC[C@@H]2[C@@]2(C)CCC(=O)C=C12 & 0.945 \Tstrut\Bstrut \\
CC(=O)[C@H]1CC[C@H]2[C@@H]3CCC4=CC(=O)CC[C@]4(C)[C@H]3CC[C@]12C & 0.944 \Tstrut\Bstrut \\
C[C@@]12[C@H](CC[C@]1(O)[C@@H]1CC[C@@H]3C[C@@H](O)CC[C@]3(C)[C@H]1C[C@H]2O)C1=CC(=O)OC1 & 0.936 \Tstrut\Bstrut \\
C[C@]12CC[C@H]3[C@@H](CCC4=CC(=O)CC[C@]34C)[C@@H]1CC[C@@H]2C(=O)CO & 0.932 \Tstrut\Bstrut \\
CC(C)(C)C(=O)OCC(=O)[C@H]1CC[C@H]2[C@@H]3CCC4=CC(=O)CC[C@]4(C)[C@H]3CC[C@]12C & 0.931 \Tstrut\Bstrut \\
C[C@]12CC(=O)C3C(CCC4=CC(=O)CC[C@]34C)C1CC[C@]2(O)C(=O)CO & 0.924 \Tstrut\Bstrut \\
CC(=O)O[C@@]1(CC[C@H]2[C@@H]3CCC4=CC(=O)CC[C@]4(C)[C@H]3CC[C@]12C)C(C)=O & 0.924 \Tstrut\Bstrut \\
CC(=O)OCC(=O)[C@H]1CC[C@H]2[C@@H]3CCC4=CC(=O)CC[C@]4(C)[C@H]3CC[C@]12C & 0.920 \Tstrut\Bstrut \\
C\textbackslash C=C1/C(=O)C[C@H]2[C@@H]3CCC4=CC(=O)CC[C@]4(C)[C@H]3CC[C@]12C & 0.919 \Tstrut\Bstrut \\
C[C@]12CC[C@H]3[C@@H](CC[C@@H]4C[C@@H](O)CC[C@]34C)[C@@]1(O)CC[C@@H]2C1=CC(=O)OC1 & 0.918 \Tstrut\Bstrut \\
CCCC(=O)O[C@@]1(CC[C@H]2[C@@H]3CCC4=CC(=O)CC[C@]4(C)[C@H]3[C@@H](O)C[C@]12C)C(=O)CO & 0.902 \Tstrut\Bstrut \\
CCC(=O)O[C@@]1(CC[C@H]2[C@@H]3CCC4=CC(=O)CC[C@]4(C)[C@H]3CC[C@]12C)C(=O)CO & 0.894 \Tstrut\Bstrut \\
CC(=O)[C@@]12OC(C)(O[C@@H]1C[C@H]1[C@@H]3CCC4=CC(=O)CC[C@]4(C)[C@H]3CC[C@]21C)c1ccccc1 & 0.880 \Tstrut\Bstrut \\
CC(=O)OCC(=O)[C@@]1(O)CCC2C3CCC4=CC(=O)CC[C@]4(C)C3[C@@H](O)C[C@]12C & 0.877 \Tstrut\Bstrut \\
CC(=O)[C@@]1(O)CCC2C3CCC4=CC(=O)CC[C@]4(C)C3CC[C@]12C & 0.876 \Tstrut\Bstrut \\
C[C@]12CCC3C(CCC4=CC(=O)CC[C@]34C)C1CC[C@@H]2O & 0.872 \Tstrut\Bstrut \\
\hline
\end{tabular}
\label{tab:fda}
\end{table}

\section{Conclusion}
In this paper, we investigate existing domain extrapolation paradigms and their limitations. To allow the method to extrapolate across combinatorially many environments, we propose a new method which complements invariant risk minimization with adaptive environments. The method is evaluated on molecule property prediction tasks and shows significant improvements over strong baselines.

\bibliography{main}

\begin{thebibliography}{12}
\providecommand{\natexlab}[1]{#1}
\providecommand{\url}[1]{\texttt{#1}}
\expandafter\ifx\csname urlstyle\endcsname\relax
  \providecommand{\doi}[1]{doi: #1}\else
  \providecommand{\doi}{doi: \begingroup \urlstyle{rm}\Url}\fi

\bibitem[Arjovsky et~al.(2019)Arjovsky, Bottou, Gulrajani, and
  Lopez-Paz]{arjovsky2019invariant}
Martin Arjovsky, L{\'e}on Bottou, Ishaan Gulrajani, and David Lopez-Paz.
\newblock Invariant risk minimization.
\newblock \emph{arXiv preprint arXiv:1907.02893}, 2019.

\bibitem[Bemis and Murcko(1996)]{bemis1996properties}
Guy~W Bemis and Mark~A Murcko.
\newblock The properties of known drugs. 1. molecular frameworks.
\newblock \emph{Journal of medicinal chemistry}, 39\penalty0 (15):\penalty0
  2887--2893, 1996.

\bibitem[Chang et~al.(2020)Chang, Zhang, Yu, and Jaakkola]{chang2020invariant}
Shiyu Chang, Yang Zhang, Mo~Yu, and Tommi~S Jaakkola.
\newblock Invariant rationalization.
\newblock \emph{arXiv preprint arXiv:2003.09772}, 2020.

\bibitem[Corsello et~al.(2017)Corsello, Bittker, Liu, Gould, McCarren,
  Hirschman, Johnston, Vrcic, Wong, Khan, et~al.]{corsello2017drug}
Steven~M Corsello, Joshua~A Bittker, Zihan Liu, Joshua Gould, Patrick McCarren,
  Jodi~E Hirschman, Stephen~E Johnston, Anita Vrcic, Bang Wong, Mariya Khan,
  et~al.
\newblock The drug repurposing hub: a next-generation drug library and
  information resource.
\newblock \emph{Nature medicine}, 23\penalty0 (4):\penalty0 405--408, 2017.

\bibitem[Ganin et~al.(2016)Ganin, Ustinova, Ajakan, Germain, Larochelle,
  Laviolette, Marchand, and Lempitsky]{ganin2016domain}
Yaroslav Ganin, Evgeniya Ustinova, Hana Ajakan, Pascal Germain, Hugo
  Larochelle, Fran{\c{c}}ois Laviolette, Mario Marchand, and Victor Lempitsky.
\newblock Domain-adversarial training of neural networks.
\newblock \emph{The Journal of Machine Learning Research}, 17\penalty0
  (1):\penalty0 2096--2030, 2016.

\bibitem[Jeon et~al.(2020)Jeon, Ko, Lee, Choi, Byun, Park, Shum, and
  Kim]{jeon2020identification}
Sangeun Jeon, Meehyun Ko, Jihye Lee, Inhee Choi, Soo~Young Byun, Soonju Park,
  David Shum, and Seungtaek Kim.
\newblock Identification of antiviral drug candidates against sars-cov-2 from
  fda-approved drugs.
\newblock \emph{bioRxiv}, 2020.

\bibitem[Long et~al.(2018)Long, Cao, Wang, and Jordan]{long2018conditional}
Mingsheng Long, Zhangjie Cao, Jianmin Wang, and Michael~I Jordan.
\newblock Conditional adversarial domain adaptation.
\newblock In \emph{Advances in Neural Information Processing Systems}, pages
  1640--1650, 2018.

\bibitem[Oord et~al.(2018)Oord, Li, and Vinyals]{oord2018representation}
Aaron van~den Oord, Yazhe Li, and Oriol Vinyals.
\newblock Representation learning with contrastive predictive coding.
\newblock \emph{arXiv preprint arXiv:1807.03748}, 2018.

\bibitem[Source(2020)]{diamond}
Diamond~Light Source.
\newblock Sars-cov-2 main protease structure and xchem fragment screen.
\newblock 2020.
\newblock
  \url{www.diamond.ac.uk/covid-19/for-scientists/Main-protease-structure-and-XChem}.

\bibitem[Yang et~al.(2019)Yang, Swanson, Jin, Coley, Eiden, Gao, Guzman-Perez,
  Hopper, Kelley, Mathea, et~al.]{yang2019analyzing}
Kevin Yang, Kyle Swanson, Wengong Jin, Connor Coley, Philipp Eiden, Hua Gao,
  Angel Guzman-Perez, Timothy Hopper, Brian Kelley, Miriam Mathea, et~al.
\newblock Analyzing learned molecular representations for property prediction.
\newblock \emph{Journal of chemical information and modeling}, 59\penalty0
  (8):\penalty0 3370--3388, 2019.

\bibitem[Zhao et~al.(2019)Zhao, Combes, Zhang, and Gordon]{zhao2019learning}
Han Zhao, Remi Tachet~des Combes, Kun Zhang, and Geoffrey~J Gordon.
\newblock On learning invariant representation for domain adaptation.
\newblock \emph{arXiv preprint arXiv:1901.09453}, 2019.

\bibitem[Zhou et~al.(2020)Zhou, Yang, Wang, Hu, Zhang, Zhang, Si, Zhu, Li,
  Huang, et~al.]{zhou2020pneumonia}
Peng Zhou, Xing-Lou Yang, Xian-Guang Wang, Ben Hu, Lei Zhang, Wei Zhang,
  Hao-Rui Si, Yan Zhu, Bei Li, Chao-Lin Huang, et~al.
\newblock A pneumonia outbreak associated with a new coronavirus of probable
  bat origin.
\newblock \emph{Nature}, 579\penalty0 (7798):\penalty0 270--273, 2020.

\end{thebibliography}
\bibliographystyle{plainnat}

\end{document}